\documentclass[prb,twocolumn]{revtex4-1}

\usepackage{graphicx}

\usepackage[utf8]{inputenc}

\usepackage{amsmath}
\usepackage{amsfonts}
\usepackage{amssymb}

\newcommand{\be}{\begin{equation}}
\newcommand{\ee}{\end{equation}}
\newcommand{\Dd}{\mathrm{d}}
\newcommand{\pa}{\partial}

\renewcommand{\doi}[1]{{\sc doi:}~\href{http://dx.doi.org/#1}{\detokenize{#1}}}

\newcommand{\jName}[1]{#1}
\newcommand{\jVol}[1]{\textbf{#1}}

\begin{document}

\title{Deriving the Schwarzschild solution from a local Newtonian limit}
% In a long title you can use \\ to force a line break at a certain location.

\author{Markus P\"ossel}
\email{poessel@hda-hd.de} % optional
%\altaffiliation[permanent address: ]{Haus der Astronom} % optional second address
% If there were a second author at the same address, we would put another 
% \author{} statement here.  Don't combine multiple authors in a single
% \author statement.
\affiliation{Haus der Astronomie and Max Planck Institute for Astronomy, K\"onigstuhl 17, 69124 Heidelberg, Germany}
% Please provide a full mailing address here.

% See the REVTeX documentation for more examples of author and affiliation lists.

\date{\today}

\begin{abstract}
The Schwarzschild metric is derived in a manner that does not require familiarity with the formalism of differential geometry beyond the ability to interpret a general spacetime metric. As such, the derivation is suitable for an undergraduate course on general relativity. The derivation uses infalling coordinates that are particularly well adapted to the situation, as well as Einstein's equation in the simple form introduced by Baez and Bunn. That version of the vacuum Einstein equations corresponds to requiring a particular local Newtonian limit: that, to first order, the deformation of a ``test ball'' of freely falling, initially-at-rest test particles is governed by the tidal forces of Newtonian gravity. 
\end{abstract}

\maketitle % title page is now complete

\section{Introduction}
The Schwarzschild solution plays a key role in teaching about general relativity: It describes the simplest version of a black hole. By Birkhoff's theorem, it more generally describes the gravitational field around any spherical mass distribution, such as the Sun in our own Solar system. As one of two particularly simple, yet physically relevant examples of a non-trivial metric (the other being the FLRW spacetime of an expanding universe), it is particularly well-suited for teaching about general techniques of ``reading'' and interpreting a spacetime metric.

Consider undergraduate courses where students are introduced to selected concepts and results from general relativity without exposing them to the full mathematical formalism. Such courses have the advantage of introducing students to one of the two great fundamental theories of 20th century physics early on (the other being quantum mechanics); they also profit from subject matter that meets with considerable interest from students.\cite{Hartle2006} Using the terminology of Christensen and Moore,\cite{Christensen2012} in the ``calculus only'' approach pioneered by Taylor and Wheeler,\cite{Taylor2001,Taylor2018} spacetime metrics are not derived, but taken as given, and the focus is on learning how to interpret a given spacetime metric. Similar presentations can be found in the first part of the ``physics first'' approach exemplified by Hartle's text book,\cite{Hartle2003} where the concepts of the metric and of geodesics are introduced early on, and their physical consequences explored, while the mathematics necessary for the Einstein equations is only introduced at a later stage. 

Whenever the approach involves an exploration of simple metrics such as the Schwarzschild solution, but stops short of the formalism required for the full tensorial form of Einstein's equations, access to a simple derivation of the Schwarzschild solution that does not make use of the advanced formalism can be a considerable advantage.

Simplified derivations of the Schwarzschild solution have a long tradition within general relativity education,\cite{Schiff1960,Harwit1973} although specific simplifications have met with criticism.\cite{Rindler1968} This article presents a derivation which requires no deeper knowledge of the formalism of differential geometry beyond an understanding of how to interpret a given spacetime metric $\Dd s^2$. The derivation avoids the criticism levelled at attempts to derive the Schwarzschild solution from the Einstein equivalence principle in combination with a Newtonian limit,\cite{Gruber1988} relying as it does on a simplified version of the vacuum Einstein equation.

More specifically, I combine the restrictions imposed by the symmetry with the simple form of Einstein's equations formulated by Baez and Bunn.\cite{BaezBunn2005} That same strategy was followed by Kassner in 2017,\cite{Kassner2017} but in this text, I use the ``infalling coordinates'' that are commonly associated with the Gullstrand-Painlev\'e form of the Schwarzschild metric,\cite{Martel2001,Visser2005,HamiltonLisle2008} not the more common Schwarzschild coordinates. That choice simplifies the argument even further. In the end, what is required is no more than the solution of an ordinary differential equation for a single function, which yields to standard methods, to obtain the desired result.

\section{Coordinates adapted to spherical symmetry and staticity}
\label{SymmetriesCoordinates}

Assume that the spacetime we are interested in is spherically symmetric and static. In general relativity, a symmetry amounts to the possibility of being able to choose coordinates that are adapted to the symmetry, at least within a restricted sub-region of the spacetime in question. That the spacetime is static is taken to mean that we can introduce a (non-unique) time coordinate ${t}$ so that our description of spacetime geometry does not depend explicitly on ${t}$, and that space and time are completely separate --- in the coordinates adapted to the symmetry, there are no ``mixed terms'' involving $\Dd {t}$ times the differential of a space coordinate in the metric. If we use ${t}$ to slice our spacetime into three-dimensional hyperplanes, each corresponding to ``space at time ${t}$,'' then each of those 3-spaces has the same spatial geometry. A mixed term would indicate that those slices of space would need to be shifted relative to another in order to identify corresponding points. The mixed term's absence indicates that in adapted coordinates, there is no need for such an extra shift. In those coordinates, we can talk about the 3-spaces as just ``space,'' without the need for specifying which of the slices we are referring to.

In the case of spherical symmetry, we can introduce spherical coordinates that are adapted to the symmetry: a radial coordinate $r$ and the usual angular coordinates $\vartheta,\varphi$, so that the spherical shell at constant $r$ has the total area $4\pi r^2$. In consequence, the part of our metric involving $\Dd\vartheta$ and $\Dd\varphi$ will have the standard form
\be
r^2(\Dd\vartheta^2+\sin^2\theta\Dd\varphi^2) \equiv r^2\Dd\Omega^2,
\ee
where the right-hand side defines $\Dd\Omega^2$, the infinitesimal solid angle corresponding to each particular combination of $\Dd\vartheta$ and $\Dd\varphi$.

The radial coordinate slices space into spherical shells, each corresponding to a particular value $r=const.$ The rotations around the origin, which are the symmetry transformations of spherical symmetry, map each of those spherical shells onto itself, and they leave all physical quantities that do not explicitly depend on $\vartheta$ or $\varphi$ invariant.

In what follows, we will use the basic structures introduced in this way --- the slices of simultaneous ${t}$, the radial directions within each slice, the angular coordinates spanning the symmetry--adapted spherical shells of area $4\pi r^2$ --- as auxiliary structures for introducing spacetime coordinates. For now, let us write down the shape that our metric has by simple virtue of the spherical symmetry, the requirement that the spacetime be static, and the adapted coordinates, namely
\be
\Dd s^2 = -c^2F(r) \Dd {t}^2 + G(r) \Dd r^2 + r^2\:\Dd\Omega^2. 
\label{StaticForm}
\ee
Students familiar with  ``reading'' a spacetime metric will immediately recognize the sign difference between the parts describing space and describing time that is characteristic for spacetime, and the speed of light $c$ that gives us the correct physical dimensions. That there is no explicit dependence on $\varphi$ and $\vartheta$ in the remaining functions $F$ and $G$ is a direct consequence of spherical symmetry. That the factor in front of $\Dd\Omega^2$ is $r^2$ is a consequence of our coordinate choice, with spherical angular coordinates so that the area of a spherical surface of constant radius $r$ is $4\pi r^2$. That there is no explicit dependence on ${t}$ is one consequence of the spacetime being static; the absence of the mixed term $\Dd {t}\cdot \Dd r$ is another. We are left with two unknown functions $F(r)$ and $G(r)$. In the following, let us call ${t}$ and $r$ the {\em static coordinates}. 
 
Note that, since $G(r)$ is as yet undefined, we have not yet chosen a specific physical meaning for the length measurements associated with our $r$ coordinate. But because of the $\Dd\Omega^2$ part, it is clear that whatever choice we make, the locally orthogonal lengths $r\cdot\Dd\vartheta$ and $r\cdot\sin\vartheta\cdot\Dd\varphi$ will have the same physical interpretation as for the length measurement corresponding to $\Dd r$.

\section{Infalling observer coordinates}
\label{Sec:InfallingObservers}

Now that we know what the radial directions are, at each moment of time ${t}$, we follow Visser\cite{Visser2005} as well as Hamilton and Lisle\cite{HamiltonLisle2008} in defining a family of radially infalling observers.  Observers in that family are in free fall along the radial direction, starting out at rest at infinity: In mapping each observer's radial progression in terms of the static coordinate time ${t}$, we adjust initial conditions, specifically: the choice of initial speed at some fixed time ${t}$, in just the right way that the radial coordinate speed goes to zero for each observer in the same way as $r\to\infty.$

It is true that talking about ``infalling'' observers already reflects our expectation that our solution should describe the spacetime of a spherically symmetric mass.  As we know from the Newtonian limit, such a mass attracts test particles in its vicinity. It should be noted, though, that all our calculations would also be compatible with the limit of no mass being present. In that case, ``infalling'' would be a misnomer, as our family of observers would merely hover in empty space at unchanging positions in $r$. 

We can imagine infinitesimal local coordinate systems associated with our observers --- think of the observer mapping out space and time by defining three orthogonal axes, and by measuring time with a co-moving clock. We assume all such little coordinate systems to be non-rotating --- otherwise, we would break spherical symmetry, since rotation would locally pick out a plane of rotation that is distinguishable from the other planes. The radial direction is a natural choice for the first space axis of those little free-falling systems. The other directions, we take to point to observers falling side by side with our coordinate-defining observer --- and to remain pointed at a specific such other observer, once the choice of direction is made.

We assume our infalling observers' clocks to be synchronised at some fixed radius value $r$. By spherical symmetry, those clocks should then be synchronised at {\em all} values of $r$. Anything else would indicate direction-dependent differences for the infalling observers and their clocks, after all. Hence, at any given static time ${t}$, all the infalling observers who are at radius value $r$ show the same proper time $T$ on the ideal clocks travelling along with them. 

Once our definition is complete, our static, spherically symmetric spacetime is filled with infalling observers from that family: Whenever we consider an event $\cal E$, there will be an observer from that family passing by at that time, at that location. 

Now, consider the coordinate speed of those infalling observers. If we position ourselves at some constant radius value $r$ and watch the falling observers fly by, then we can express both their proper time rate and their coordinate speed in the $r$ direction in terms of $r$ and ${t}$. We can combine the two pieces of information to obtain the rate of change in radial position $r$ with proper time $T$ for those infalling observers. But since the initial conditions for those observers are the same, and since our spacetime is, by assumption, static, the resulting function can only depend on $r$, and not explicitly on ${t}$. Let us rescale that function with the speed of light to make it dimensionless, give it an overall minus sign to make it positive for infalling particles, and call it $\beta(r)$,
\be
\beta(r)\equiv -\frac{1}{c}\frac{\Dd r}{\Dd T}(r).
\label{betaDefinition}
\ee

Recall from section \ref{SymmetriesCoordinates} that we also still have the freedom to decide on the physical meaning of $r$. We make the choice of making $\Dd r$ the physical length measured by one of our infalling observers at the relevant location in spacetime, at constant time $T$. Via our angular coordinates, that implies that length measurements orthogonal to the radial direction, $r\cdot\Dd\vartheta$ and $r\cdot\sin\vartheta\:\Dd\varphi$ inherit the same physical interpretation.

As a next step, we transform our metric (\ref{StaticForm}) from the static form into the form appropriate for our coordinate choice $r$ and $T$. We do so by writing the static time coordinate as a function ${t}(T,r)$ in terms of infalling observer time and radius value. In consequence,
\be
\Dd {t} = \frac{\pa{t}}{\pa T}\cdot\Dd T+ \frac{\pa {t}}{\pa r}\cdot\Dd r,
\ee
and our new metric now has the form
\begin{align}
 \Dd s^2 = {} & -c^2 F(r)\left(\frac{\pa t}{\pa T}\right)^2\Dd T^2 \nonumber \\[0.2em]
 & -2c^2F(r)\left(\frac{\pa t}{\pa T}\right)\left(\frac{\pa t}{\pa r}\right)\Dd T\:\Dd r \nonumber \\[0.2em]
 & +\left[G(r)-c^2F(r)\left(\frac{\pa t}{\pa r}\right)^2\right]\Dd r^2+r^2\:\Dd\Omega^2. %\\[0.5em]
% = {} & -c^2\bar{F}(r,T)\Dd T^2 -2\bar{H}(r,T)\Dd T\:\Dd r \nonumber \\[0.2em]
%& +\bar{G}(r,T)\Dd r^2+r^2\:\Dd\Omega^2 
 \end{align}
At face value, this looks like we are moving the wrong way, away from simplification, since we now have more functions, and they depend on two variables instead of one.

But in fact, this new formulation paves the way for an even simpler form of the metric. Consider a specific event, which happens at given radius value $r$. In a small region around that event, we will introduce a new coordinate $\bar{r}$ to parametrize the radial direction. We want this coordinate to be co-moving with our infalling observers at $r$; each such observer then has a position $\bar{r}=const.$ that does not change over time. 

Key to our next step is that we {\em know} the metric for the local length and time measurements made by any one of our free-falling observers. By Einstein's equivalence principle, the metric is that of special relativity. Locally, namely whenever tidal effects can be neglected, spacetime geometry for any non-rotating observer in free fall is indistinguishable from Minkowski spacetime as described by a local inertial system.

Since we have chosen both the time coordinate $T$ and the physical meaning of the radial coordinate $r$ so as to conform with the measurements of the local infalling observer, the transformation between $\bar{r}$ and $r$ is particularly simple: It has the form of a Galilei transformation
\be
\Dd\bar{r}= \Dd r + \beta(r)c\:\Dd T.
\label{barRshift}
\ee
In that way, as it should be by definition, radial coordinate differences at constant $T$ are the same in both systems, while for an observer at constant $\bar{r},$ with $\Dd \bar{r}=0$, the relation between $\Dd r$ and $\Dd T$ is consistent with the definition of the function $\beta(r)$ in (\ref{betaDefinition}).

Are you surprised that this is not a Lorentz transformation, as one might expect from special relativity? Don't be. We are not transforming from one local inertial coordinate system to another. The $T$ is already the time coordinate of the infalling observers, so both coordinate systems have the same definition of simultaneity, and time dilation plays no role in this particular transformation. Also, we have chosen $r$ intervals to correspond to length measurements of the infalling observers, so there is no Lorentz contraction, either. It is the consequence of these special choices that gives the relation (\ref{barRshift}) its simple form.

Last but not least, when we analyse specifically an infinitesimal neighbourhood of the point $r,\vartheta,\varphi$, let us make the choice that directly at our point of interest, we make $\bar{r}$ coincide with $r$. Since before, we had only fixed the differential $\Dd \bar{r}$, we do have the remaining freedom of choosing a constant offset for $\bar{r}$ that yields the desired result.

By Einstein's equivalence principle, the metric in terms of the locally co-moving coordinates $T,\bar{r},\vartheta,\varphi$ is the spherical-coordinate version of the Minkowski metric,
\be
\Dd s^2 = -c^2\Dd T^2 + \Dd\bar{r}^2 + \bar{r}^2\Dd\Omega.
\ee
This version can, of course, be obtained by taking the more familiar Cartesian-coordinate version
\be
\Dd s^2=-c^2\Dd T^2 + \Dd X^2 + \Dd Y^2 + \Dd Z^2,
\label{CartesianMinkowski}
\ee
applying the definition of Cartesian coordinates $X,Y,Z$ in terms of spherical coordinates $\bar{r},\vartheta,\varphi$
\be
x= \bar{r}\:\sin\vartheta\:\cos\varphi, \;\;
y= \bar{r}\:\sin\vartheta\:\sin\varphi, \;\;
z= \bar{r}\:\cos\vartheta,
\ee
to express $\Dd X, \Dd Y, \Dd Z$ in terms of $\Dd \bar{r}, \Dd\vartheta, \Dd\varphi$, and substitute the result into (\ref{CartesianMinkowski}).

By noting that we have chosen $\bar{r}$ so that, at the specific spacetime event where we are evaluating the metric, $\bar{r}=r$, while, for small radial coordinate shifts around that location, we have the relation (\ref{barRshift}), we can now write down the same metric in the coordinates $T, r, \vartheta,\varphi$, namely as
\be
\Dd s^2 = -c^2\left[
1-\beta(r)^2
\right] \Dd T^2+2c\beta(r)\Dd r\:\Dd T
+\Dd r^2+r^2\Dd\Omega^2.
\label{preMetric}
\ee
Since we can repeat that local procedure at any event in our spacetime, this result is our general form of the metric, for all values of $r$. This, then is the promised simplification: By exploiting the symmetries of our solutions as well as the properties of infalling observers, we have reduced our metric to a simple form with no more than one unknown function of one variable, namely $\beta(r)$.

So far, what I have presented is no more than a long-form version of the initial steps of the derivation given by Visser in his heuristic derivation of the Schwarzschild metric.\cite{Visser2005} In the next section, we will deviate from Visser's derivation.

\section{$\beta(r)$ from tidal deformations}
\label{TidalSection}

In the previous section, we had exploited symmetries and Einstein's equivalence principle. In order to determine $\beta(r)$, we need to bring in additional information, namely the Einstein equations, which link the matter content with the geometry of spacetime. For our solution, we only aim to describe the spacetime metric outside whatever spherically-symmetric matter distribution resides in (or around) the center of our spherical symmetry. That amounts to applying the {\em vacuum Einstein equations}.

More specifically, we use a particularly simple and intuitive form of the vacuum Einstein equations, which can be found in a seminal article by Baez and Bunn:\cite{BaezBunn2005} Consider a locally flat free-fall system around a specific event $\cal E$, with a time coordinate $\tau$, local proper time, where the event we are studying corresponds to $\tau=0$. In that system, describe a small sphere of freely floating test particles, which we shall call a {\em test ball}. The particles need to be at rest relative to each other at $\tau=0$. Let the volume of the test ball be $V(\tau)$. Then the vacuum version of Einstein's equations states that
\be
\left.\frac{\Dd^2 V}{\Dd\tau^2}\right|_{\tau=0} = 0.
\label{EinsteinVacuum}
\ee
In words: If there is no matter or energy inside, the volume of such a test ball remains constant in the first order (those were our initial conditions) and the second order (by eq.~[\ref{EinsteinVacuum}]). 

If you are familiar with Wheeler's brief summary of Einstein's equations, ``spacetime grips mass, telling it how to move'' and ``mass grips spacetime, telling it how to curve'',\cite{Wheeler1990} you will immediately recognise that this is a specific way for the structure of spacetime telling the test ball particles how to move. The calculation later in this section provides the second part: It will amount to using (\ref{EinsteinVacuum}) to determine the structure of spacetime, namely the still missing function $\beta(r)$, and that is the way for mass, in this case: for the absence of mass, to tell spacetime how to curve.

Note that equation (\ref{EinsteinVacuum}) also holds true in Newtonian gravity. So in a way, this version of Einstein's equation can be seen as a second-order extension of the usual Einstein equivalence principle: Ordinarily, the equivalence principle is a statement about physics in the absence of tidal forces. Equation (\ref{EinsteinVacuum}) adds to this that the lowest-order correction for tidal forces in a freely falling reference frame is that specified by Newtonian gravity. This makes sense, since by going into a free-fall frame, and restricting our attention to a small spacetime region, we have automatically created a weak-gravity situation. In such a situation, tidal corrections are approximately the same as those described by Newton. This argument can serve as a heuristic justification of (\ref{EinsteinVacuum}).

In 2017, Kassner made use of the Baez-Bunn form of Einstein's vacuum equation to derive the Schwarzschild solution, starting from what we have encountered as the static form of the metric (\ref{StaticForm}).\cite{Kassner2017} We follow the same general recipe, but using the infalling coordinates introduced in section \ref{Sec:InfallingObservers}, which makes our derivation even simpler.

\begin{figure}[htbp]
\begin{center}
\includegraphics[width=0.5\linewidth]{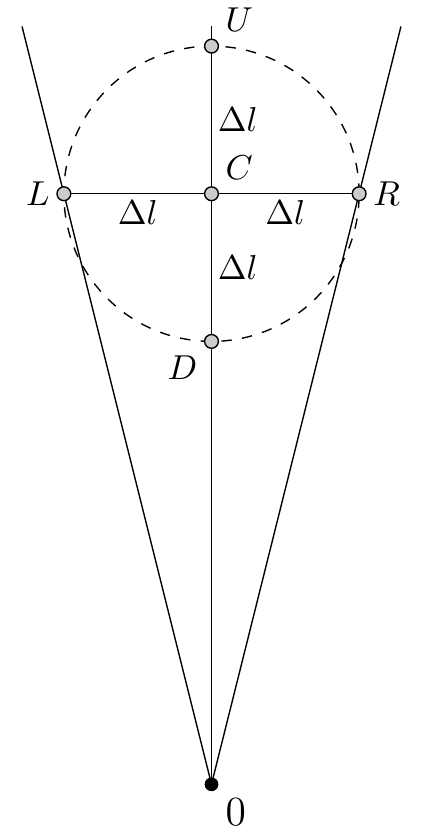}
\caption{Five test particles in our spherically-symmetric spacetime}
\label{TestParticlesOutside}
\end{center}
\end{figure}

Consider five test particles in a small region of space. Let the motion of each be the same as for the local representative from our coordinate-defining family of infalling observers. We take the central particle $C$ to be at radial coordinate value $r=R$ at the time of the snapshot shown in Fig.~\ref{TestParticlesOutside}. The other four are offset relative to the central particle: As described in the local inertial system that is co-moving with the central particle, one of the particles is shifted by $\Delta l$ upwards in the radial direction, another downward, while two of the particles are offset orthogonally by the same distance.
The $\Delta l$ is meant to be infinitesimally small, so while Fig.~\ref{TestParticlesOutside} is of course showing a rather large $\Delta l$ so as to display the geometry of the situation more clearly, we will in the following only keep terms linear in $\Delta l$. 

Consider a generic particle, which moves as if it were part of our coordinate-defining family of infalling observers, and which at the time $T_0$ is at $r=r_0$. By a Taylor expansion, that particle's subsequent movement is given by
\be
r(T) = r_0 + \frac{\Dd r}{\Dd T}(T_0) \cdot \Delta T +\frac12 \frac{\Dd^2 r}{\Dd T^2}(T_0) \cdot \Delta T^2
\label{TaylorREvo}
\ee
where $\Delta T\equiv T-T_0$. We know from (\ref{betaDefinition}) that the derivative in the linear term can be expressed in terms of $\beta(r)$; by the same token,
\be
\frac{\Dd^2 r}{\Dd T^2} = -c\frac{\Dd\beta}{\Dd T}=-c\beta' \frac{\Dd r}{\Dd T} = c^2\beta\cdot\beta',
\ee
where the prime denotes differentiation of $\beta$ with respect to its argument. Since, in the following, the product of $\beta$ and its first derivative will occur quite often, let us introduce the abbreviation
\be
B(r) \equiv \beta(r)\cdot\beta'(r).
\label{BigBDefinition}
\ee
With these results, can rewrite the Taylor expansion (\ref{TaylorREvo}) as 
\be
r(T) = r_0 -c\beta(r_0)\cdot\Delta T + \frac12 c^2B(r_0)\cdot\Delta T^2.
\label{RadialOrbitTime}
\ee
In order to find $r_C(T)$ for our central particle, we simply insert $r_0=R$ into that expression. If, on the other hand, we want to write down the time evolution for particles $U$ and $D$, let us denote it by $r_{U,D}(T)$, we need to evaluate the expression (\ref{RadialOrbitTime}) at the initial location $r_0=R\pm\Delta l$. Since $\Delta l$ is small, we can make a Taylor expansion of $\beta(r)$ and its derivative around $r=R$, and neglect everything beyond the terms linear in $\Delta l$. The result is
\begin{multline}
r_{U,D}(T)=R \pm\Delta l-c\left[
\beta(R)\pm\beta'(R)\Delta l
\right]\Delta T \\[0.2em]
+\frac{c^2}{2}\big[
B(R)\pm B'(R)\Delta l
\big]\Delta T^2
\end{multline}
In consequence, the distance between the upper and lower particle, $d_{\parallel}(T)\equiv r_U(T)-r_D(T),$ changes over time as
\be
d_{\parallel}(T) =  2\Delta l\left[
1-c\beta'(R)\Delta T+\frac12c^2 B'(R)\Delta T^2
\right].
\label{dParallel}
\ee
Next, let us look at how the distance between the particles $L$ and $R$ changes over time. The initial radial coordinate value for each of the particles is
\be
r(T_0) = \sqrt{R^2+\Delta l^2}=R\left[1+\frac12\left(\frac{\Delta l}{R}\right)^2\right]\approx R,
\ee
that is, equal to $R,$ as long as we neglect any terms that are higher than linear in $\Delta l$. In consequence, $r_{L,R}(t)$ is the same function as for our central particle, given by eq.~(\ref{RadialOrbitTime}) with $r_0=R$. The transversal (in Fig.~\ref{TestParticlesOutside}: horizontal) distance $d_{\perp}(T)$ between the particles $L$ and $R$ changes in proportion to the radius value,
\begin{align}
d_{\perp}(T) &= 2\Delta l\cdot\frac{r_{L}(T)}{R} \nonumber \\
                 &=2\Delta \left[1-\frac{c\beta(R)}{R}\Delta T+\frac{c^2}{2}\frac{B(R)}{R}\Delta T^2\right].
                \label{dPerp}
\end{align}
With these preparations, consider the vacuum Einstein equation (\ref{EinsteinVacuum}) for the volume of a test ball. Initially, our particles $C, U, D, L, R$ define a circle, which is deformed to an ellipse. By demanding rotational symmetry around the radial direction, we can construct the associated ellipsoid, which is initially a spherical surface. That ellipsoid has one axis in radial direction, whose length is $d_{\parallel}(T)$, and two axes that are transversal and each have the length  $d_{\perp}(t)$. But that ellipsoid is not quite yet the test ball we need. After all, the particles of the test ball need to be at rest initially, at time $T_0$, in the co-moving system defined by the central particle $C$. Our defining particles are not, as the terms linear in $\Delta T$ in both (\ref{dParallel}) and (\ref{dPerp}) show, where the coefficients of $\Delta T$ correspond to the particles' initial velocities. 

In order to define our test ball, we need to consider particles at the same location, undergoing the same acceleration, but which are initially at rest relative to the central particle $C$. 

We could go back to the drawing board, back to Fig.~\ref{TestParticlesOutside}, make a more general Ansatz that includes initial velocities which measure the divergence of the motion of our test ball particles from that of the infalling-observer particles, and repeat our calculation while including those additional velocity terms. But there is a short-cut. The only consequence of those additional velocity terms will be to change the terms linear in $\Delta T$ in equations (\ref{dParallel}) and (\ref{dPerp}). And we already know the end result: We will choose the additional terms so as to cancel the terms linear in $\Delta T$ in the current versions of (\ref{dParallel}) and (\ref{dPerp}). But by that reasoning, we can skip the explicit steps in between, and write down the final result right away. The time evolution of the radial-direction diameter of our test ball, let us call it $L_{\parallel}(T)$, must be the same as $d_{\parallel}(T)$, but without the term linear in $\Delta T$. Likewise, the time evolution $L_{\perp}(T)$ of the two transversal diameters must be equal to $d_{\perp}(T)$, but again without the term linear in $\Delta T$. The result is
\begin{align}
L_{\parallel}(T)  &=  2\Delta l \left[1+\frac12c^2B'(R)\Delta T^2\right] \\
L_{\perp}(T) &= 2\Delta l \left[1+\frac{c^2}{2}\frac{B(R)}{R}\Delta T^2\right].
\end{align}
Thus, our test ball volume is
\begin{align}
V(T) &= \frac{\pi}{6}L_{\parallel}(T) L_{\perp}^2(T) \\
       &= \left.\frac{4\pi}{3}\Delta l^3\left[1+{c^2}\left( \frac{B(r)}{r} + \frac{B'(r)}{2}\right)\Delta T^2\right]\right|_{r=R}
\end{align}
For the second time derivative of $V(T)$ to vanish at the time $T=T_0$, we must have
\be
\frac{B(r)}{r} + \frac{B'(r)}{2}= 0
\label{VolumeConditionR}
\ee
for all values of $r$. This is readily solved by the standard method of separation of variables: We can rewrite (\ref{VolumeConditionR}) as
\be
\frac{\Dd B}{B} = -2\frac{\Dd r}{r},
\ee
which is readily integrated to give
\be
\ln(B) = -\ln(r^{2}) + const.  \;\; \Rightarrow \;\; \ln(Br^2) = C',
\ee
with a constant $C'$, which upon taking the exponential gives us
\be
Br^2= C,
\label{BSolution}
\ee
with a constant $C$. Note that the constant $C$ can be negative --- there is no reason the constant $C'$ needs to be real; only our eventual function $B(r)$ needs to be that, and it is clear that (\ref{BSolution}) satisfies the differential equation
(\ref{VolumeConditionR}) for any constant $C$, positive, zero, or negative. By (\ref{BigBDefinition}), the solution (\ref{BSolution}) corresponds to the differential equation
\be
\beta(r)\beta'(r) = \frac{C}{r^2}
\ee
for our function $\beta$; with another separation of variables, we can re-write this as 
\be
\beta\cdot\Dd\beta=C\frac{\Dd r}{r^2}.
\ee
Both sides are readily integrated up; we can solve the result for $\beta(r)$ and obtain
\be
\beta(r) = \sqrt{
-\frac{2C}{r} +2D
},
\ee
where $D$ is the second integration constant, and where we have chosen the proper sign, since we know that $\beta(r)>0$. That brings us to the last step: The requirement that, for large values of $r$, the description provided by our solution should correspond to the results from Newtonian gravity. First of all, we note that our initial condition for the infalling observers, which had those observers start out at zero speed at infinity, means that we must choose $D=0$. Then, as we would expect, $\beta(r)$ for large values of $r$ becomes very small, corresponding to small speeds. But at slow speeds, time and length intervals as measured by the infalling observer will become arbitrarily close to time and length intervals as measured by an observer at rest in our static coordinate system at constant $r$, using the static time coordinate ${t}$. As is usual, we identify these coordinates with those of an approximately Newtonian description. In that description, the radial velocity is
\be
v(r) = \sqrt{\frac{2GM}{r}},
\ee
which follows directly from energy conservation for the sum of each observer's kinetic and Newtonian-gravitational potential energy. This fixes the remaining integration constant as
\be
C = -\frac{GM}{c^2},
\ee
and the final form of our function $\beta(r)$ becomes
\be
\beta(r) = \sqrt{\frac{2GM}{rc^2}}.
\ee
Inserting this result in (\ref{preMetric}), we obtain the metric
\be
\Dd s^2 = -c^2\left[
1-\frac{2GM}{rc^2}
\right]\Dd T^2+2\sqrt{\frac{2GM}{r}}\Dd r\:\Dd T+\Dd r^2+r^2\Dd\Omega^2.
\label{GPMetric}
\ee
This is known as the Gullstrand-Painlev\'e version of the Schwarzschild metric.\cite{Martel2001,Visser2005,HamiltonLisle2008} A  last transformation step brings us back to the traditional Schwarzschild form. Recall our discussion in sec.~\ref{SymmetriesCoordinates}, leading up to the explicitly static form (\ref{StaticForm}) of the metric? The main difference between our current form and the static version is the mixed term containing $\Dd r\:\Dd T$ in (\ref{GPMetric}). Everything else already has the required shape. Inserting the Ansatz
\be
\Dd T = \Dd t + \xi(r) \Dd r
\ee
into the metric (\ref{GPMetric}), it is straightforward to see that the mixed term vanishes iff our transformation is
\be
\Dd T = \Dd t +\frac{\sqrt{2GM/r}}{c^2\left(1-\frac{2GM}{rc^2}\right)}\Dd r.
\label{TtTrafo}
\ee
% PN9 41
Substitute this into (\ref{GPMetric}), and the result is the familiar form of the Schwarzschild metric in Schwarzschild's original coordinates $t,r,\vartheta,\varphi$, 
\be
\Dd s^2 = -c^2\left(1-\frac{2GM}{c^2 r}
\right)\Dd t^2 + \frac{\Dd r^2}{\left(1-\frac{2GM}{c^2 r}
\right)} + r^2\Dd\Omega^2.
\ee
%PN9 43

\section{Conclusion}
Using coordinates adapted to the symmetries, we were able to write down the spherically symmetric, static spacetime metric. On this basis, and using the family of infalling observers that is characteristic for the Gullstrand-Painlev\'e solution, we wrote down the metric in the form (\ref{preMetric}), with a single unknown function $\beta(r)$. From the simplified form (\ref{EinsteinVacuum}) of the vacuum Einstein equations, as applied to a test ball in free fall alongside one of our family of observers, we were able to determine $\beta(r)$, up to two integration constants. By using the Einstein equation, we escape the restrictions imposed on simplified derivations by Gruber et al.\cite{Gruber1988} 

From the initial condition for our infalling observers, as well as from the Newtonian limit at large distances from our center of symmetry, we were able to fix the values of the two intergration constants. Our derivation does not require knowledge of advanced mathematical concepts beyond the ability to properly interpret a given metric line element $\Dd s^2$. Even our analysis of tidal effects proceeds via a simple second-order Taylor expansion, leading to differential equations for $\beta(r)$ that are readily solved using two applications of the method of separation of variables. 

What is new about the derivation presented here is the combination of the Baez-Bunn equations with the infalling coordinates typical for the Gullstrand-Painlev\'e form of the metric --- this combination is what, in the end, makes our derivation particularly simple. In turn, this simplicity is what should make the derivation particularly useful in the context of teaching general relativity in an undergraduate setting.

The derivation proceeds close to the physics, and gives ample opportunity to discuss interesting properties of Einstein's theory of gravity. Students who are presented with this derivation, either as a demonstration or as a (guided) exercise, will come to understand the way that symmetries determine the form of a metric, the deductions that can be made from Einstein's equivalence principle, and last but not least that we need to go beyond the equivalence principle, and consider tidal forces, to completely define our solution.

\section*{Acknowledgements}

I would like to thank Thomas M\"uller for helpful comments on an earlier version of this text.

\end{document}